\documentclass{appolb}
\usepackage{graphicx}

\newcommand{\be}{\begin{equation}}
\newcommand{\ee}{\end{equation}}
\newcommand{\ba}{\begin{eqnarray}}
\newcommand{\ea}{\end{eqnarray}}
\newcommand{\bd}{\begin{displaymath}}
\newcommand{\ed}{\end{displaymath}}
\newcommand{\bea}{\begin{eqnarray}}
\newcommand{\eea}{\end{eqnarray}}

\begin{document}
\title{EMBEDDING A CRITICAL POINT IN A HADRON TO QUARK-GLUON CROSSOVER EQUATION OF STATE%
\thanks{Presented at the 29th International Conference on Ultrarelativistic Nucleus-Nucleus Collisions}%
}
\author{Joseph Kapusta, Tom Welle
\address{School of Physics \& Astronomy, University of Minnesota, Minneapolis, Minnesota 55455, USA}
\\[3mm]
{Christopher Plumberg 
\address{Illinois Center for Advanced Studies of the Universe, Department of Physics, University of Illinois at Urbana-Champaign, Urbana, Illinois 61801, USA}
}
}
\maketitle
\begin{abstract}
It is shown how to embed a critical point in a smooth background equation of state so as to yield the critical exponents and critical amplitude ratios expected of a transition in the same universality class as the liquid--gas phase transition and the 3D Ising model.  There are only two independent critical exponents; the relations 
$\alpha + 2\beta + \gamma = 2$ and $\beta (\delta - 1) = \gamma$ arise automatically, as does a new relation between the two critical amplitudes.  The resulting equation of state has parameters which may be inferred by hydrodynamic modeling of heavy ion collisions in the Beam Energy Scan II at the Relativistic Heavy Ion Collider.
\end{abstract}
  
\section{Introduction}
Many model calculations \cite{stephanov,CPOD} predict the existence of a critical point in the QCD phase diagram at a value of the chemical potential where current lattice simulations \cite{Bazavov2020,Borsanyi2020} are unreliable.  How to combine or merge a critical equation of state with a smooth background is a long-standing problem in statistical physics with no unique solution.  Our goal is to construct an equation of state in the same universality class as the liquid--gas phase transition and the 3D Ising model. It should have parameters which may be inferred by hydrodynamic modeling of heavy ion collisions in the Beam Energy Scan II at the Relativistic Heavy Ion Collider or in experiments at other accelerators.  Such an equation of state is also needed for modeling neutron star mergers.

\section{Construction}

Motivated by S-shaped curves in first order phase transitions and by the cubic equation we define the functions
\bd
Q_{\pm}(T,\mu) = \left\{  \left[ (\Delta^2(T))^{2} + r^{2}(T,\mu) \right]^{1/2} \pm r(T,\mu)  \right\}^k
\ed 
\bd
r(T,\mu) = \frac{\mu^4 - \mu_x^4(T)}{\mu^4 + \mu_x^4(T)} \;\;\; {\rm and}  \;\;\; \Delta^2(T) \sim d_{\pm}|T/T_c - 1|^p \;\;  {\rm for} \;\;T \rightarrow T_c^{\pm}
\ed
where $\mu_x(T)$ is the chemical potential along the coexistence curve.  There are only two exponents $k$ and $p$.  The smooth background pressure $P_{BG}$ is modified via multiplication by a function $R$ as
\bd
P(T,\mu) = P_{BG}(T,\mu) R(T,\mu)
\ed
When ${\small T \ge T_c}$
\bd
R(T,\mu) = 1 - a(T)\left( \sqrt{ \Delta^4 + 1} + 1 \right)^k - a(T)\left( \sqrt{ \Delta^4 + 1} - 1 \right)^k + a(T)(Q_+ + Q_-)
\ed
When ${\small T \le T_c}$ and ${\small \mu \le \mu_x(T)}$
\bd
R_h = 1 + a(T) Q_-(T,\mu) - a(T) \left( \sqrt{ \Delta^4 + 1} + 1 \right)^k
\ed
and when ${\small \mu \ge \mu_x(T)}$
\bd
R_q = 1 + a(T) Q_+(T,\mu) - a(T) \left( \sqrt{ \Delta^4 + 1} + 1 \right)^k
\ed
Here $a(T)$ is a smooth monotonically decreasing function of $T$.  The critical exponents are derived to be $\alpha = 2-kp$, $\beta = (k-1)p$, $\gamma = (2-k)p$, and $\delta = 1/(k-1)$.  These satisfy the well known relations $\alpha + 2\beta + \gamma = 2$ and $\beta (\delta - 1) = \gamma$.  This construction predicts a previously unknown relation between universal ratios of critical amplitudes
\bd
\left(\frac{c_+}{c_-}\right)^{2-k} = 4 \left(\frac{\chi_-}{\chi_+}\right)^k
\ed
where the $c_\pm$ are the critical amplitudes for the heat capacity and the $\chi_\pm$ are the critical amplitudes for the susceptibility.  This relation is in agreement with published results \cite{Guida,Hasenbusch1,Hasenbusch2}. 
 
\section{Results}

It remains to specify the function $\mu_x(T)$.  In order to have an inverted U-shaped coexistence curve in the $T-n$ plane, as seen in the argon and carbon dioxide liquid-gas phase transitions, the function $\mu_x(T)$ is determined by ${\small R(T,\mu_x(T)) n_{BG}(T,\mu_x(T)) = n_c}$.  For purposes of illustration, we take the background equation of state from \cite{matchingpaper}.  This background equation of state smoothly interpolates from a hadronic gas with excluded volume interactions at low energy density to an interacting quark-gluon gas at high energy density.  Specific parameter choices and more details may be found in \cite{us}.

In this construction there are nine parameters.  The two independent critical exponents and the one independent ratio of critical amplitudes are universal.  The numerical values of $T_c$ and $\mu_c$ are obviously not universal and are of utmost interest.  Finally there are four parameters which determine the extent of the critical region in the temperature versus chemical potential plane.  For example, do the true critical exponents begin to manifest themselves within 10\%, 1\%, or even 0.1\% of the critical point?  That is not universal.
\begin{figure}[h]
\centerline{%
\includegraphics[width=10.5cm]{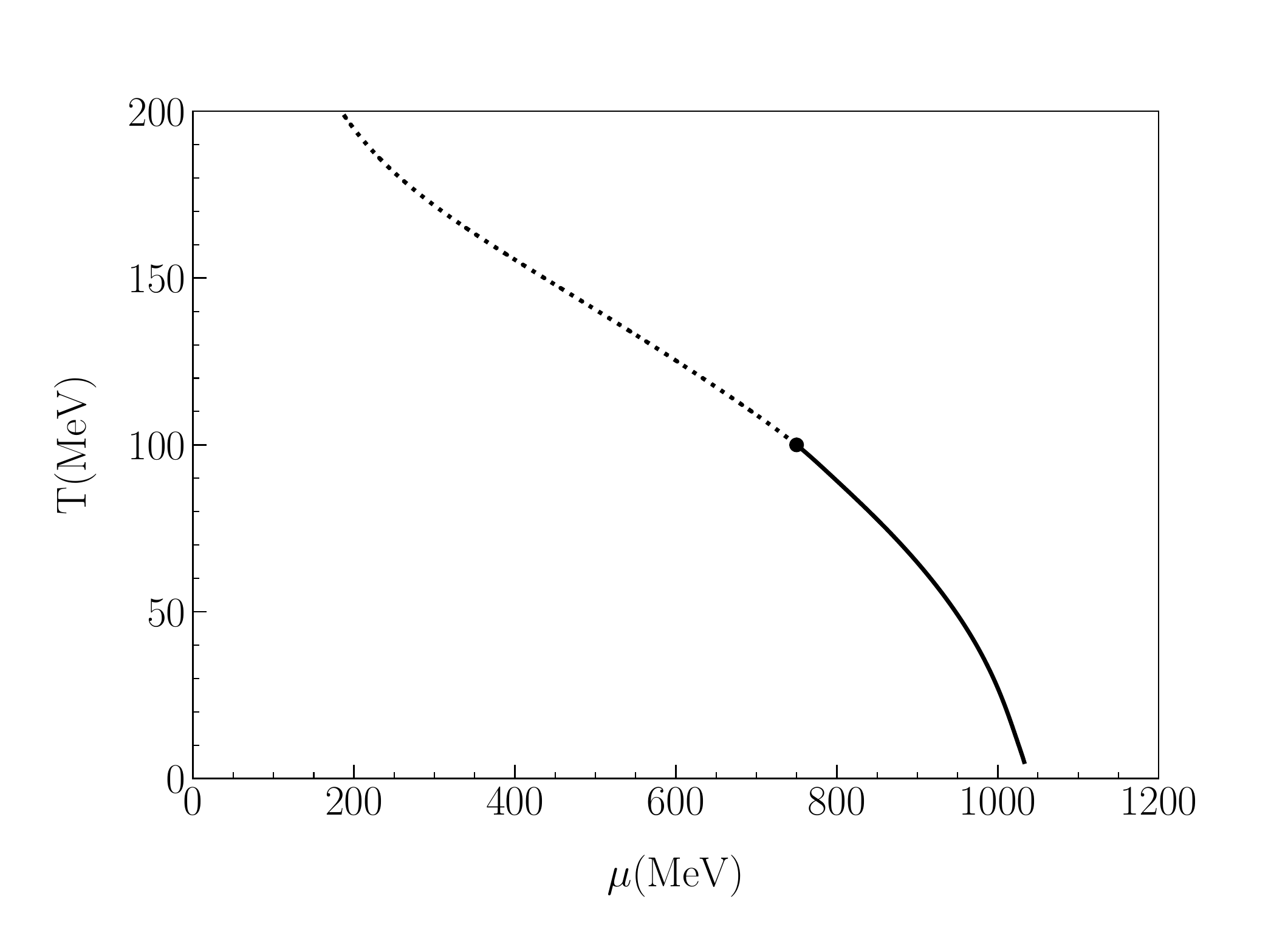}}
\caption{The coexistence curve is indicated by the solid line with a critical point chosen to lie at $T_c = 100$ MeV and $\mu_c = 750$ MeV.  The function $\mu_x(T)$ must be defined for all $T$ without any cusps or singularities.}
\end{figure}
The resulting coexistence curve is shown in Fig. 1. 
The function $R$ is shown in Fig. 2.  This function really exhibits the critical behavior embedded in the smooth background equation of state.  The pressure versus density curve is shown in Fig. 3.  The dotted points represent the Maxwell construction between the coexisting state.  It is clearly symmetric about the critical point.
\begin{figure}[h]
\centerline{%
\includegraphics[width=10.5cm]{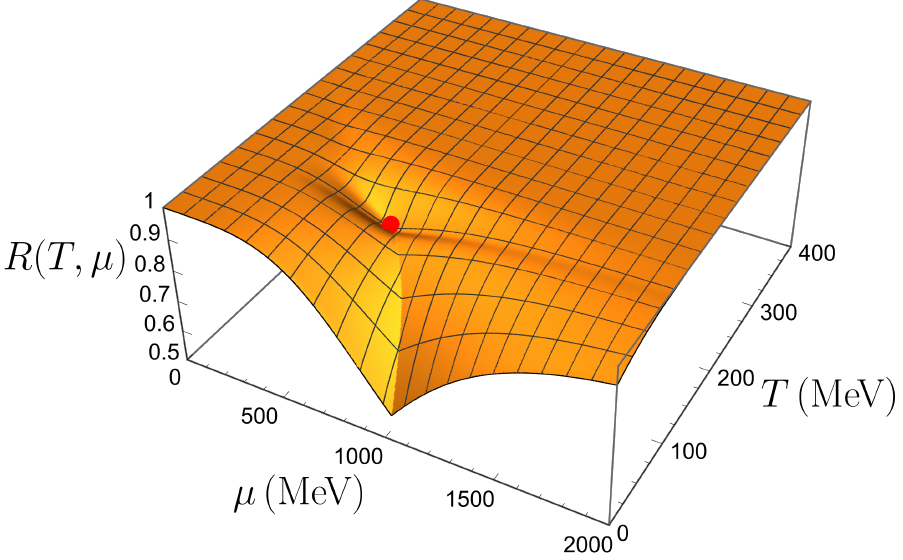}}
\caption{Plot of the function $R(T,\mu)$.  The dot indicates the location of the critical point.}
\end{figure}
\begin{figure}[h]
\centerline{%
\includegraphics[width=10.5cm]{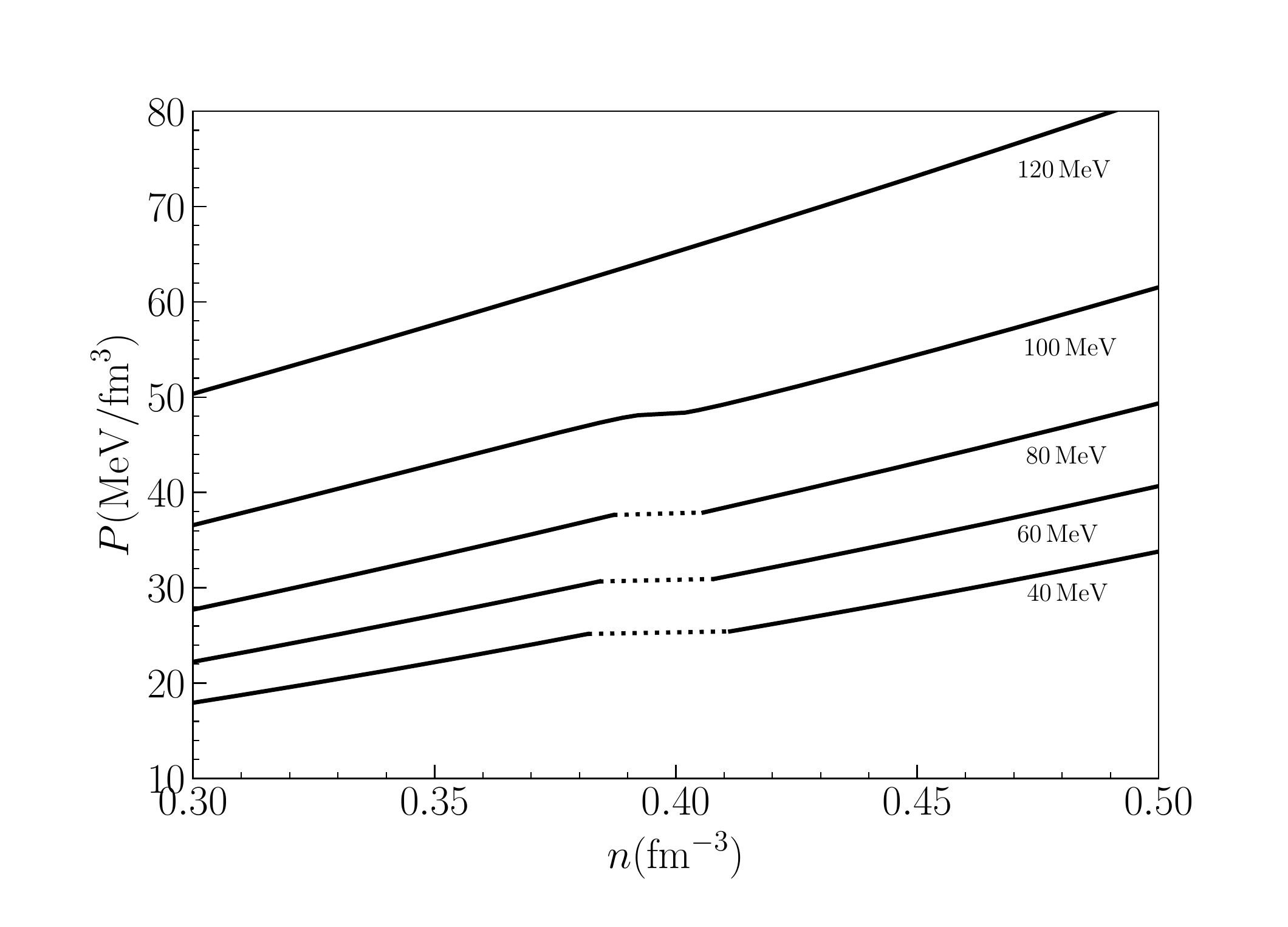}}
\caption{Isotherms of pressure versus density.  The critical isotherm at $T = 100$ MeV appears to show a discontinuity only because of the large critical exponent
$\delta$.} 
\end{figure}

\newpage

\section{Conclusion}
Lattice QCD simulations have shown unequivocally that the transition from hadrons to quarks and gluons is a crossover when the baryon chemical potential is zero or small \cite{Aoki2006,Bhattacharya2014}.  Many model calculations predict the existence of a critical point at a value of the chemical potential where current lattice simulations are unreliable.  We show how to embed a critical point in a smooth background equation of state so as to yield the critical exponents and critical amplitude ratios expected of a transition in the same universality class as the liquid--gas phase transition and the 3D Ising model.  There are only two independent critical exponents; the usual relations among them is automatically satisfied.  There arises a new relation between the two critical amplitudes.  The resulting equation of state has parameters which may be inferred by hydrodynamic modeling of heavy ion collisions in the Beam Energy Scan II at the Relativistic Heavy Ion Collider or in experiments at other accelerators.  

\section*{Acknowledgments}
The work of JK and TW was supported by the U.S. DOE Grant No. DE-FG02-87ER40328.  The work of CP was supported by the U.S. DOE Grant No. DE-SC0020633

\end{document}